# Wafer-scale solution-derived molecular gate dielectrics for low-voltage graphene electronics

*Vinod K. Sangwan,[1] Deep Jariwala,[1] Ken Everaerts,[2] Julian J. McMorrow,[1] Jianting He,[1] Matthew Grayson,[3] Lincoln J. Lauhon,[1] Tobin J. Marks,[1,2]\* and Mark C. Hersam[1,2]\**

[1]Department of Materials Science and Engineering, Northwestern University, Evanston, Illinois 60208, USA

[2]Department of Chemistry, Northwestern University, Evanston, Illinois 60208, USA

[3]Department of Electrical Engineering and Computer Science, Northwestern University, Evanston, Illinois 60208, USA

*Authors to whom correspondence should be addressed.
Electronic mail: t-marks@northwestern.edu and m-hersam@northwestern.edu

**Abstract**

Graphene field-effect transistors are integrated with solution-processed multilayer hybrid organic-inorganic self-assembled nanodielectrics (SANDs). The resulting devices exhibit low-operating voltage (2 V), negligible hysteresis, current saturation with intrinsic gain > 1.0 in vacuum (pressure < $2 \times 10^{-5}$ Torr), and overall improved performance compared to control devices on conventional $SiO_2$ gate dielectrics. Statistical analysis of the field-effect mobility and residual carrier concentration demonstrate high spatial uniformity of the dielectric interfacial properties and graphene transistor characteristics over full 3 inch wafers. This work thus establishes SANDs as an effective platform for large-area, high-performance graphene electronics.

**Manuscript**

Graphene, an atomically-thin, two-dimensional honeycomb lattice of $sp^2$-bonded carbon, has emerged as a promising candidate material for next-generation electronics due to its high intrinsic field-effect mobility (~200,000 $cm^2/Vs$), mechanical robustness, and chemical stability.[1-6] However, in field-effect transistors, the mobility of graphene can be compromised by charged impurities,[7,8] microscopic ripples,[9] and remote interface phonons in the gate dielectric layer.[1,7,10] While graphene field-effect transistors (G-FETs) fabricated in a suspended geometry[5] or on mechanically exfoliated boron nitride flakes[11] have shown ultra-high carrier mobility, these approaches have only been demonstrated on isolated devices and pose significant challenges to large-scale integration.

Alternatively, mobility in graphene can be enhanced on high-κ gate dielectrics as a result of reduced Coulomb interactions between the carriers and charged impurities in the dielectrics and/or adsorbates.[1,7,12] Furthermore, high-κ gate dielectrics enable aggressive channel scaling for low-power operation and higher switching speed.[2,13-17] Nevertheless, the conventional growth techniques for high-κ dielectrics, such as atomic layer deposition (ALD) and physical vapor deposition, require high temperatures and/or capital-intensive high vacuum processes.[18-20] Therefore, significant efforts have been devoted to developing alternative growth routes for high-κ gate dielectrics. Solution-based growth strategies (*e.g.*, sol-gel and molecular self-assembly) enable low-temperature, low-cost syntheses of high-κ dielectrics for large-area circuitry as well as for unconventional electronics on flexible and transparent polymeric substrates.[21-24] Early efforts to integrate graphene with solution-processable self-assembled monolayers (SAMs) showed promise in lowering operating voltages and improving carrier mobility versus conventional SiO$_2$ dielectrics, although the wafer-scale uniformity of these organic dielectrics has not yet been established.[25]

In this Letter, we utilize a novel class of solution-processable gate dielectric, self-assembled nanodielectrics (SANDs), to realize high-performance G-FETs at the wafer-scale. SANDs consist of alternating layers of high-κ inorganic oxides and a highly polarizable phosphonic acid π-electron stilbazolium group (PAE = 4-[[4-[bis(2-hydroxyethyl)-amino]phenyl]diazenyl]-1-[4-(diethoxyphosphoryl) benzyl]pyridinium bromide).[23,26-30] The multilayer SAND structure has previously allowed 1 nm precise thickness control with ultra-low current leakage (<10$^{-7}$ A/cm$^2$) and high capacitance (>1 µF/cm$^2$).[30] SAND variants have been developed to optimize transistor performance for different semiconducting materials (*e.g.*, organics,[28] nanowires,[31,32] and carbon nanotubes[29,30,33]) under a range of processing conditions (*e.g.*, solution-phase[26,28,30] and vapor-phase[27,29]). In particular, SANDs based on phosphonic acid linkages are attractive due to facile fabrication in ambient conditions at relatively low temperatures (150 ºC) on metal oxides including alumina, zirconia, and hafnia.[28,30] Here, we integrate large-area graphene (2" × 2") grown by chemical vapor deposition (CVD) with solution-processable hafnia-based SAND (Hf-SAND)[30] grown on 3" Si wafers. In addition, the Hf-SAND surface, which is terminated by a layer of sol-gel derived HfO$_x$, is functionalized with phosphonic acid SAMs to achieve controlled doping of the resulting G-FETs. When compared with control G-FETs on conventional SiO$_2$/Si substrates, the SAND-based devices exhibit desirable attributes such as reduced operating voltage, negligible hysteresis, and current saturation with wafer-scale uniformity.

Fig. 1a shows a 2-probe bottom-contact G-FET on four layers of Hf-SAND (Hf-SAND-4). Hf-SAND-4 was fabricated on a degenerately n-doped Si wafer (with native oxide) following reported procedures[30]. Hf-SAND-4 shows a low leakage current density (< $10^{-6}$ A/cm$^2$) and high capacitance of 485 nF/cm$^2$ up to an electric field of 1.3 MV/cm (effective oxide thickness = 6.84 nm, assuming overall κ = 9.2;[30] see Section S1 in Supplementary Material for SAND characterization). Source-drain electrodes (2 nm Ti/ 70 nm Au) were patterned by photolithography. To prevent electrical shorting of probe contacts to the substrate, a 10 nm thick ALD-grown Al$_2$O$_3$ layer (100 °C) was deposited underneath the metal pads prior to thermal evaporation and lift-off. Then, as-purchased CVD graphene (Graphene Supermarket) was transferred from its Cu foil onto the patterned electrodes on Hf-SAND using a modified literature procedure[34-36] (see Section S2 in Supplementary Material). The process was optimized to minimize wrinkles and tears in the transferred graphene. Bottom-contact geometry graphene channels were patterned photolithographically and isolated by reactive ion etching (RIE) (Figs. 1c, d). Finally, the devices were cleaned in resist stripper (Remover PG, MicroChem Inc.).

Raman spectra of a completed G-FET on Hf-SAND (Fig. 1e) shows that the defect D peak (~1350 cm$^{-1}$) intensity is less than 0.1 times that of the G peak (~1587 cm$^{-1}$). Furthermore, the 2D : G peak intensity ratio (2 − 3) and the 2D peak shape suggests that this transfer process preserves the high quality of the as-grown single-layer CVD graphene.[37,38] Note that the graphene G peak on Hf-SAND is blue-shifted versus that in undoped graphene (G ~ 1580 cm$^{-1}$), indicating that the graphene on Hf-SAND is doped under ambient conditions. As will be seen later, G-FET electrical measurements in ambient conditions reveal strong p-doping (> 6 × 10$^{12}$ cm$^{-2}$), consistent with the reported G peak up-shifts in electrostatically doped graphene.[39,40]

The G-FETs on Hf-SAND were characterized in ambient as well as under vacuum (pressure < 2 × 10$^{-5}$ Torr) using Keithley 2400 SourceMeters. Note that proof-of-concept G-FETs were also fabricated on zirconia-based SAND (Zr-SAND,[28] Section S3 in Supplementary

Material), but Hf-SAND was selected for more systematic studies due to the superior dielectric performance.[30] Fig. 2a shows transfer characteristics ($I_d$-$V_g$) of a device with channel length $L$ = 70 μm and channel width $W$ = 150 μm. The devices in ambient show substantial gate hysteresis as well as significant p-doping (gate voltage ($V_g$) at the Dirac point $V_{Dirac}$ > 2.0 V). Such p-type behavior in ambient is commonly observed in G-FETs on conventional $SiO_2$ dielectrics. Under vacuum, the G-FETs on Hf-SAND show negligible hysteresis and approximately symmetric ambipolar behavior with the Dirac point near 0 V, suggesting that Hf-SAND itself induces negligible doping in graphene. The field-effect mobility ($\mu$) and other relevant device metrics in 2-probe G-FETs are extracted by applying a widely used model.[36,41] In particular, the total device resistance ($R$) is given by eq. 1:

$$R = 2R_c + \frac{L/W}{e\mu\sqrt{n_o^2 + n^2}} \qquad (1)$$

where $R_c$ is the contact resistance, $n_0$ is the residual carrier concentration near the Dirac point, $e$ is the fundamental unit of charge, and $n$ is the carrier density induced by the gate electrode. The linear dispersion of graphene at low energies results in a carrier density dependent quantum capacitance that dominates the overall capacitance in G-FETs, especially in devices with a large geometric capacitance ($C_{Hf-SAND}$).[42] Thus, the quantum capacitance is taken into account to more accurately estimate the carrier concentration. Assuming that the Fermi level $E_F \gg k_B T$ (where $k_B$ is the Boltzmann constant and $T$ is temperature), the relationship between $V_g$ and $n$ is given by eq. 2:

$$V_g - V_{Dirac} = \frac{en}{C_{Hf-SAND}} + \frac{\hbar v_F \sqrt{\pi n}}{e} \qquad (2)$$

where $\hbar$ is Planck's constant divided by $2\pi$, and $v_F$ is the Fermi velocity of graphene ($10^6$ m/s).[36] The fitting of the data in Fig. 2b (fitting parameters: $\mu$, $n_0$, $V_{Dirac}$, and $R_c$) results in $\mu = 4,330$ cm$^2$/Vs, $n_0 = 3.7 \times 10^{11}$ cm$^{-2}$, $V_{Dirac} = -0.12$ V, and $R_c = 168$ Ω. Electron and hole mobilities can also be extracted separately by fitting the left and the right sides of the $R$-$V_g$ curve, respectively. Since the electron and hole mobilities differ by less than 5% in G-FETs on Hf-SAND, we consider only hole mobility in the following discussion. It is likely that the field-effect mobilities of the present G-FETs are limited by the intrinsic CVD graphene quality.

To demonstrate wafer-scale uniformity of CVD graphene with Hf-SAND, a large number of devices (total = 251) were characterized over a 50 × 50 mm$^2$ area on a 3" wafer, and a high-mobility subset of these devices (total = 39) were characterized within a 3 × 4 mm$^2$ subarea. Figs. 3a-d show histograms (dashed) of the device parameters extracted from fitting the Dirac voltage $V_{Dirac}$, doping level $n_{Dirac}$, residual carrier concentration $n_0$, and field-effect mobility $\mu$. The solid histograms in Fig. 3a-d show the same metrics extracted from the subarea. The overall mean mobility value of $\mu_m = 2,280$ cm$^2$/Vs was observed with a standard deviation of $\Delta\mu = 880$ cm$^2$/Vs. The residual carrier concentration $n_0$ would be zero for disorder-free graphene and has been shown to correlate well with the density of charged impurities estimated from a rigorous self-consistent theory of electronic transport in graphene.[7,36] In the present work, 95% of the devices have $n_0$ in the range 3 - 8 × 10$^{11}$ cm$^{-2}$ with the mean $n_0 = 5.3 \times 10^{11}$ cm$^{-1}$, and 95% of the devices have $V_{Dirac}$ within a ~1 V range (−0.25 V to +0.72 V) with the mean $V_{Dirac} = 0.17$ V ($n_{Dirac}$ ~ 5.1 × 10$^{11}$ cm$^{-2}$).

We now examine the high mobility subarea of $3 \times 4$ mm$^2$. Note that the solid histogram of $V_{Dirac}$, $n_{Dirac}$, and $n_0$ (see Fig. 3a-c) from the devices within the $3 \times 4$ mm$^2$ subarea are significantly more homogeneous than the overall histogram. In particular, 93% of the $V_{Dirac}$ distribution lies within a 0.2 V range (0.0 V to +0.2 V) suggesting that this fabrication method is capable of generating uniform, high quality devices over areas large enough for integrated circuits. The solid histogram of field-effect mobility $\mu$ (see Fig. 3d) tends to be somewhat broader than the narrow distribution of $V_{Dirac}$ and $n_0$, since $V_{Dirac}$ and $n_0$ only depend on the dielectric environment of graphene (e.g. trap charges in SAND), whereas the distribution in $\mu$ is also affected by the spatial variation in the defects in CVD graphene. The mean field-effect mobility $\mu_m = 2,840$ cm$^2$/Vs is approximately 25% larger than mean $\mu_m$ of the devices over the 3" wafer. Thus, the higher uniformity SAND at a 5 mm length-scale bodes well for eventual wafer-scale uniformity of SAND *via* more stringently controlled processing conditions.

To gain more insight into the high performance of G-FETs on Hf-SAND, we consider control graphene transistors on a 300 nm SiO$_2$ gate dielectric (see Section S4 in Supplementary Material for details). A clear advantage of the Hf-SAND gate dielectric is the improved transconductance ($g_m = \partial I_d / \partial V_g$). Thus, the channel geometry normalized intrinsic transconductance of the G-FET on Hf-SAND in Fig. 2a ($L.g_m/W \sim 137$ µS at $V_d = 0.1$ V) is more than 60 times larger than that of the best control device ($g_m \sim 2.2$ µS at $V_d = 0.1$ V). The mean mobility of G-FETs on Hf-SAND is also found to be more than 2 times greater than that of the control G-FETs on SiO$_2$ (830 cm$^2$/Vs), and the average mobility of the best 5% Hf-SAND devices is also more than 2 times higher than the best 5% of the control devices. The mean residual carrier concentration in the control devices is $n_0 \sim 1.2 \times 10^{12}$ cm$^{-2}$. Thus, $n_0$ in G-FETs on Hf-SAND is less than half that in the control devices fabricated under identical conditions.

According to a self-consistent charge scattering model, the mobility of high quality graphene flakes on Hf-SAND with the observed $n_0$ and κ values is expected to be as high as µ ~ 2.5 m$^2$/Vs, representing an upper mobility limit achievable with this method. This discrepancy again suggests that the mobility in the best G-FETs on Hf-SAND is limited by the intrinsic defects/wrinkles in the transferred CVD graphene. The higher mobility and reduced residual charge density in the Hf-SAND G-FETs likely arises from the reduced trapped charge density at the Hf-SAND surface compared to control SiO$_2$ devices and/or improved Coulomb screening from the higher effective κ of Hf-SAND.[1,7,8,10] Previously, the favorable interfacial properties of SAND enabled superior transistor performance in organic semiconductors[28] and carbon nanotube thin films[29,30,33] *versus* conventional oxide dielectrics. In addition, the interfacial properties of Hf-SAND can be further tuned by functionalizing the surface with phosphonic acid SAM dopant layers to achieve the controlled doping of semiconductors for low-power complementary metal-oxide-semiconductor (CMOS) integrated circuits.[25,43,44] Here, G-FETs on wafer-scale Hf-SAND overcoated with octadecylphosphonic acid are significantly n-doped ($n_{Dirac}$ ~ 2.5 × 10$^{12}$ cm$^{-2}$) without significant degradation in mobility (see Section S5 in Supplementary Material).

The large electron mobility of graphene suggests potential applications in RF analog electronics.[45] However, current saturation in G-FETs is essential to realize large gains in amplifying circuits. The unique graphene gapless electronic structure impedes realizing full current saturation (*i.e.*, output conductance $g_{ds} = \partial I_d / \partial V_d \sim 0$) over a large range of drain biases.[41,46] Instead, ambipolar G-FETs (and ambipolar carbon nanotube FETs) show a characteristic kinked behavior.[41,47] Ultra-thin high-κ dielectrics have been extensively explored in top-gated graphene transistors to improve the current saturation behavior.[13,16,41,48] In Fig. 4a, a three-dimensional plot of output characteristics of a bottom-gated G-FET on Hf-SAND shows the expected kinked behavior. This device shows desirable current saturation (minimum channel width-normalized $g_{ds}$ ~ 7 µS/µm at $V_d$ = 2.5 V, Fig. 4b). The theoretical maximum gain achievable in an amplifier is represented by the intrinsic gain figure-of-merit ($g_m/g_{ds}$), where a gain of more than unity is essential for operational circuits. Here, even the smallest channel graphene/Hf-SAND devices ($L$ ~ 2 µm) show desirable current saturation (minimum $g_{ds}$ ~ 14 µS/µm at maximum $g_{ds}$ = 35 µS/µm) and an intrinsic gain exceeding 2.

In conclusion, we have demonstrated wafer-scale integration of CVD graphene with ambient solution-processed high-κ molecular gate dielectrics. The devices exhibit high field-effect mobility, negligible hysteresis, low operating voltage (±2 V), and desirable current saturation. The Hf-SAND G-FET devices also show higher mobility and a lower density of residual carriers near the charge neutrality point versus control devices on $SiO_2$. Spatially, 93% of the Hf-SAND G-FETs within an area of approximately 3 × 4 $mm^2$ showed Dirac voltages within a 0.24 V range. The high spatial uniformity of the Hf-SAND G-FET devices at the wafer-scale suggests new opportunities for large-area graphene electronics.

**Acknowledgements:** This work was supported by the Northwestern University Materials Research Science and Engineering Center (NSF DMR-1121262) and the Office of Naval Research MURI Program (ONR N00014-11-1-0690) in addition to the Northwestern International Institute for Nanotechnology. J.J.M. acknowledges support from a NASA Space Technology Research Fellowship (NSTRF, #NNX12AM44H), and J.H. acknowledges support from a Meister Summer Research Grant. The authors thank B. Myers and I. S. Kim for assistance with lithography and Raman spectroscopy, respectively. The authors also acknowledge R. Divan and L. Ocola of the Center for Nanoscale Materials at Argonne National Laboratory for assistance with clean room fabrication. Use of the Center for Nanoscale Materials was supported by the U. S. Department of Energy, Office of Science, Office of Basic Energy Sciences, under Contract No. DE-AC02-06CH11357. This research also utilized the NUFAB cleanroom facility at Northwestern University and the NUANCE Center at Northwestern University, which is supported by the NSF-MRSEC, Keck Foundation, and the State of Illinois.

**Figures:**

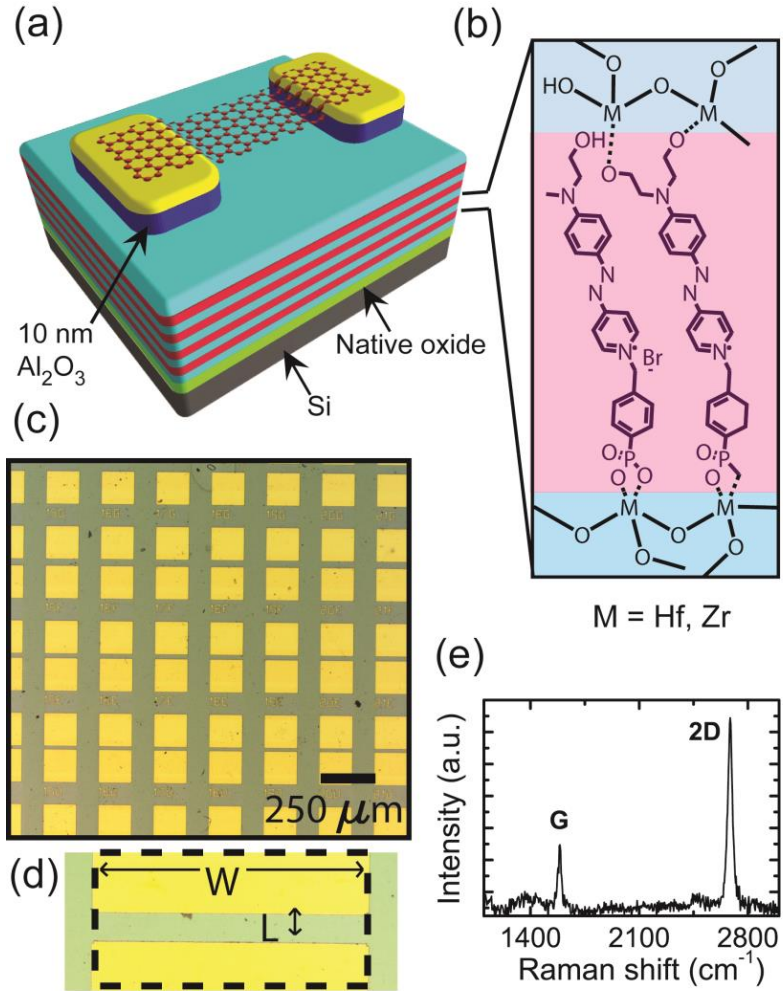

FIG. 1. (a) Schematic of a graphene field-effect transistor (G-FET) on a 4-layer self-assembled nanodielectric (SAND) on a Si substrate. (b) Chemical structure of the PAE molecule (length ~ 1.5 nm) sandwiched between sol-gel derived metal oxide layers ($MO_x$, where M = Hf or Zr). (c) Optical micrograph of several 2-probe G-FETs on 4-layer Hf-SAND. The black scale bar = 250 μm. (d) Magnified optical micrograph of a G-FET where the dashed black line outlines the patterned CVD graphene. (e) Representative Raman spectrum of CVD graphene on Hf-SAND in ambient using 514 nm laser excitation with a power of ~250 μW (100x objective). The spectral acquisition time was limited to 90 sec, and the Raman light was dispersed by a 600 grooves/mm holographic grating.

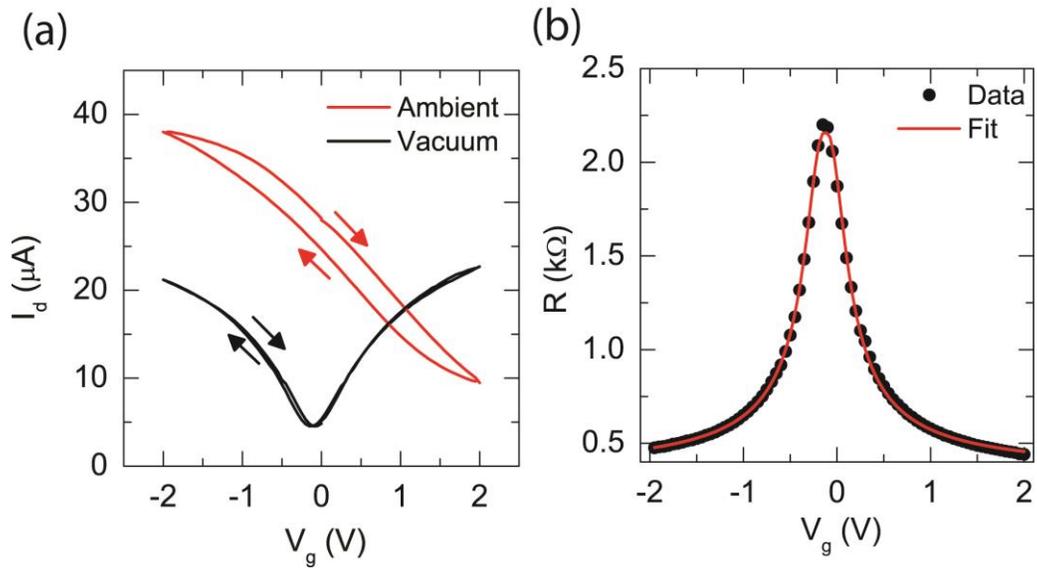

FIG. 2. (a) Transfer characteristics of a graphene/ Hf-SAND G-FET ($L = 70$ μm, $W = 150$ μm) at $V_d = 10$ mV in ambient as well as under vacuum ($< 2 \times 10^{-5}$ Torr). (b) Resistance versus gate voltage of the same data as in (a) fit to eq. 1.

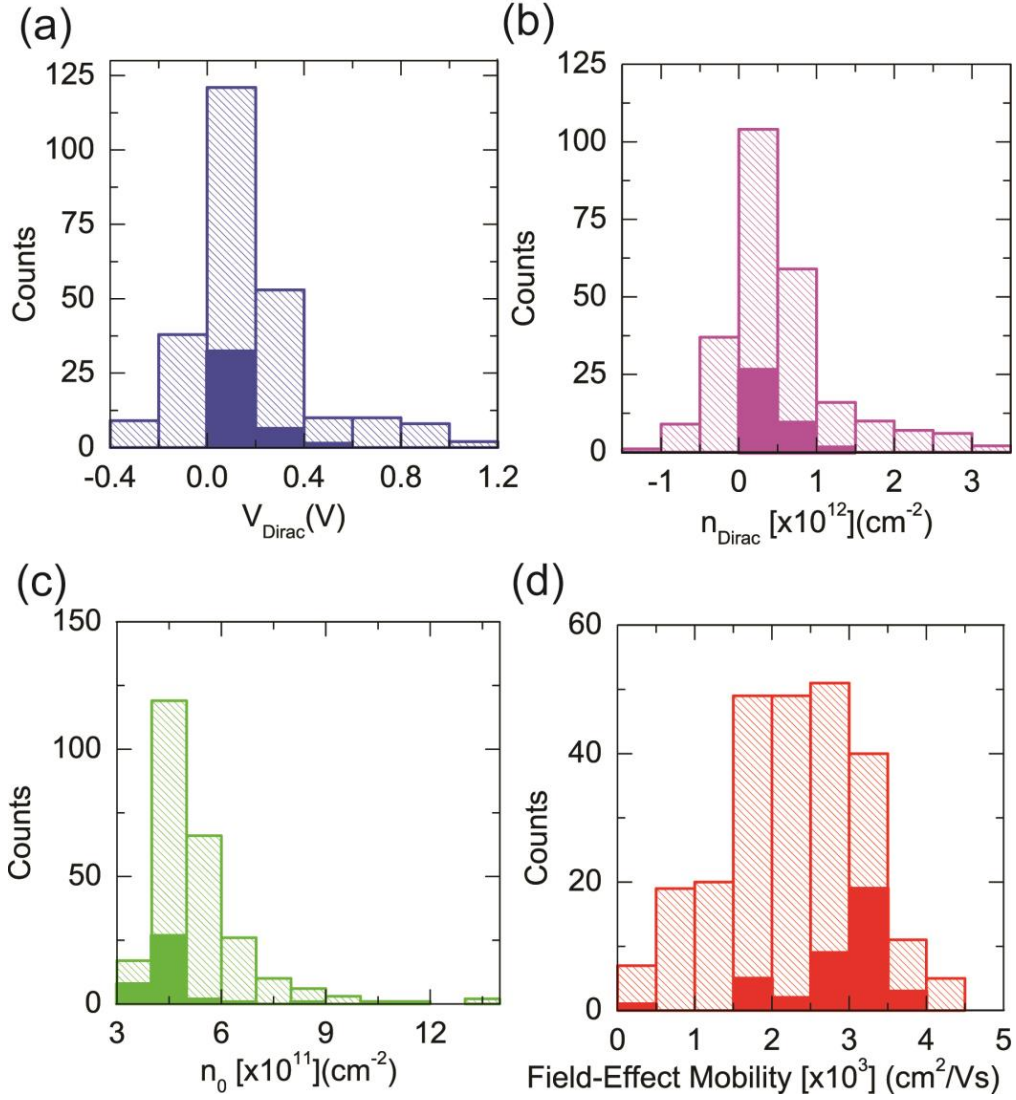

FIG. 3. Histograms (dashed) of (a) the Dirac voltage $V_{Dirac}$, (b) the doping level $n_{Dirac}$, (c) the residual carrier concentration $n_0$, and (d) the field-effect mobility µ of all 251 graphene G-FETs measured on 4-layer Hf-SAND on a 3" Si wafer. The solid histograms in each of the sub-figures are from 39 devices measured within a subarea of $3 \times 4$ mm$^2$.

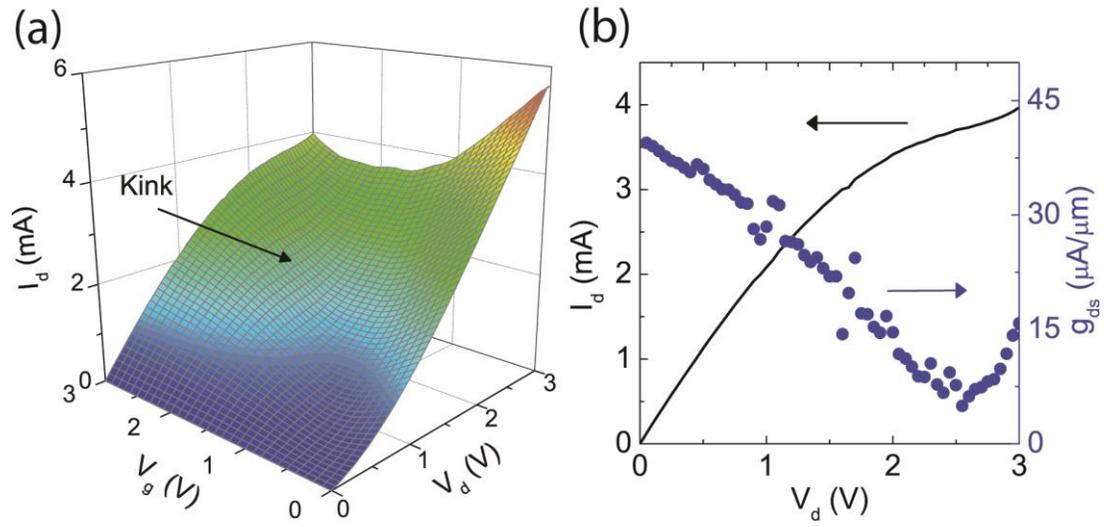

FIG. 4. (a) Three-dimensional plot of the output characteristics of a G-FET ($L = 50$ μm, $W = 150$ μm) on 4-layer Hf-SAND. (b) Output characteristics and output conductance of the same device at $V_g = 3$ V.

**References**

See supplementary material at [URL will be inserted by AIP] for characterization of SAND, description of the graphene transfer process, characteristics of G-FETs on Zr-SAND, characteristics of control G-FETs, and controlled SAM doping on Hf-SAND.

# Supplementary Material

# Wafer-scale solution-derived molecular gate dielectrics for low-voltage graphene electronics

*Vinod K. Sangwan,[1] Deep Jariwala,[1] Ken Everaerts,[2] Julian J. McMorrow,[1] Jianting He,[1] Matthew Grayson,[3] Lincoln J. Lauhon,[1] Tobin J. Marks,[1,2]\* and Mark C. Hersam[1,2]\**

[1]Department of Materials Science and Engineering, Northwestern University, Evanston, Illinois 60208, USA

[2]Department of Chemistry, Northwestern University, Evanston, Illinois 60208, USA

[3]Department of Electrical Engineering and Computer Science, Northwestern University, Evanston, Illinois 60208, USA

\*Authors to whom correspondence should be addressed.
Electronic mail: t-marks@northwestern.edu and m-hersam@northwestern.edu

**Section S1: SAND Characteristics**

Metal-insulator-semiconductor (MIS) capacitors were fabricated on a 4-layer Hf-SAND (Hf-SAND-4) wafer by evaporating 200 nm thick Au through shadow masks (200 μm x 200 μm). Leakage current-voltage (I-V) and capacitance-voltage (C-V) characteristics were measured in ambient using a 4200-SCS Parameter Analyzer. Figs. S1a and b show typical I-V and C-V characteristics.

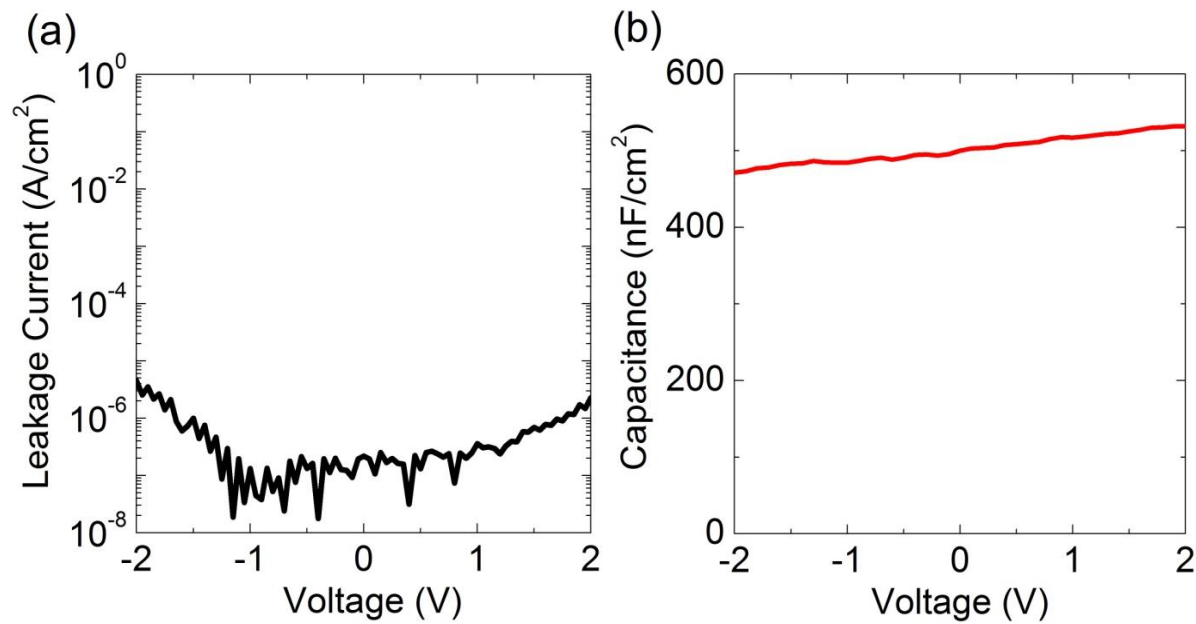

Figure S1. (a) Representative leakage current-voltage characteristics of a typical MIS capacitor on a 4-layer Hf-SAND wafer. (b) Representative capacitance-voltage characteristics (10 kHz) of the same MIS capacitor.

**Section S2: Transfer of Graphene on a SAND Wafer**

Cu foils (2 inch x 2 inch) with graphene grown by chemical vapor deposition (CVD) on both sides were purchased from Graphene Supermarket, Inc. To transfer graphene, the following procedure was adapted from Ref[1-3]. First, a Cu foil was coated with PMMA by spinning the PMMA solution (A4, MicroChem Corp.) at 2500 rpm for 2 min and then baked at 40 °C for 1 hr. The graphene on the back side of the foil was then etched via reactive ion etching (RIE) in $O_2$ plasma for 40 sec (power = 50 mW, flow-rate = 20 sccm, pressure = 100 mTorr). The foil was left floating on Marble's reagent (20 g $CuSO_4$ + 100 mL HCl + 200 mL $H_2O$) contained in a large beaker for 10 min with the PMMA-coated side facing up. Then, the foil was gently raised and the bottom side was slowly rubbed with a cotton swab to remove unetched parts of the back-side

CVD graphene. The Cu foil was left floating in Marble's reagent until all of the Cu dissolved. Then, the floating PMMA-coated graphene film was gently removed by scooping it with a large glass dish. It was then successively transferred and rinsed in three large beakers filled with Nanopure DI water (18.2 MΩ-cm). Subsequently, Hf-SAND[4] wafers were dipped into water, and the floating graphene/PMMA film was gently scooped in the middle of the wafer. The coated wafer was left in vacuum (~ 50 mbar) overnight to completely dry off the water. The wafer was baked at 190 °C (ramp rate = 4 °C per min) for 1 hr to release the stress built in the PMMA film during transfer and improve adhesion of the graphene/PMMA stack to the Hf-SAND substrate. Finally, PMMA was dissolved in acetone for 30 min at 35 °C. The graphene film was then etched into device channels by photolithography and RIE. Finally, G-FETs were cleaned in Remover PG at 80 °C (MicroChem Corp.).

**Section S3: Characteristics of G-FETs on Zirconia-Based SAND**

In order to generalize the integration of graphene with SANDs, G-FETs were also fabricated on 4-layers of zirconia-based SAND (Zr-SAND).[5] In this case, 2-probe devices were fabricated by transferring mechanically exfoliated single-layer graphene from a $SiO_2$/Si substrate onto a Zr-SAND substrate. Fig. S2 shows representative gate voltage dependent conductance of the resulting device. C-V characteristics of Zr-SAND showed a large variation in capacitance with bias due to formation of depletion region in these more lightly n-doped Si substrates. A conservative estimate (assuming the lowest capacitance) yielded field-effect mobilities approaching 5,000 $cm^2$/Vs in ambient conditions.

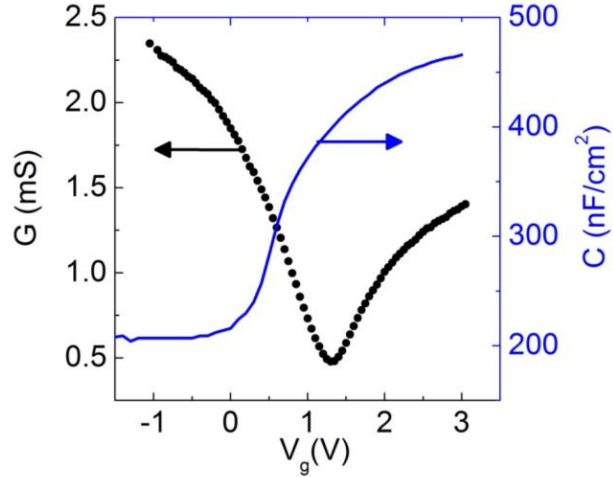

Figure S2. Transfer characteristics of a G-FET ($L = 3.2$ µm, $W = 3.5$ µm) on 4-layer Zr-SAND measured in ambient conditions. The right axis shows C-V characteristics (10 kHz) of the MIS capacitor. $V_g$ for C-V means the bias applied to the top Au electrode of the MIS capacitor.

**Section S4: Characteristics of Control G-FETs**

Control G-FETs were fabricated on a 1 inch x 1 inch 300 nm $SiO_2$/Si substrate following the procedure identical to that used for G-FETs on Hf-SAND. The devices were measured in vacuum (pressure $< 2 \times 10^{-5}$ Torr). Fig. S3a shows a representative $I_d$–$V_g$ characteristic of a device ($L = 15$ µm, $W = 20$ µm) with large operating voltage ($\pm 50$ V) compared to a G-FET on Hf-SAND ($\pm 2$ V) from Fig. 2a of the main text. Control devices show a Dirac point near 0 V in vacuum, suggesting low doping. Field-effect mobility was extracted by fitting the device resistance to equation 1 of the main text. Note that the quantum capacitance was not included here due to the low geometrical capacitance of 300 nm $SiO_2$ (~11 nF/cm$^2$). Histograms of device metrics (Figs. S3b-d) show the uniformity of transferred CVD graphene. However, the residual charge density in the control devices was found to be rather large ($n_0 \sim 1.2 \times 10^{12}$ cm$^{-2}$), possibly due to contamination from the transfer process and photolithography.

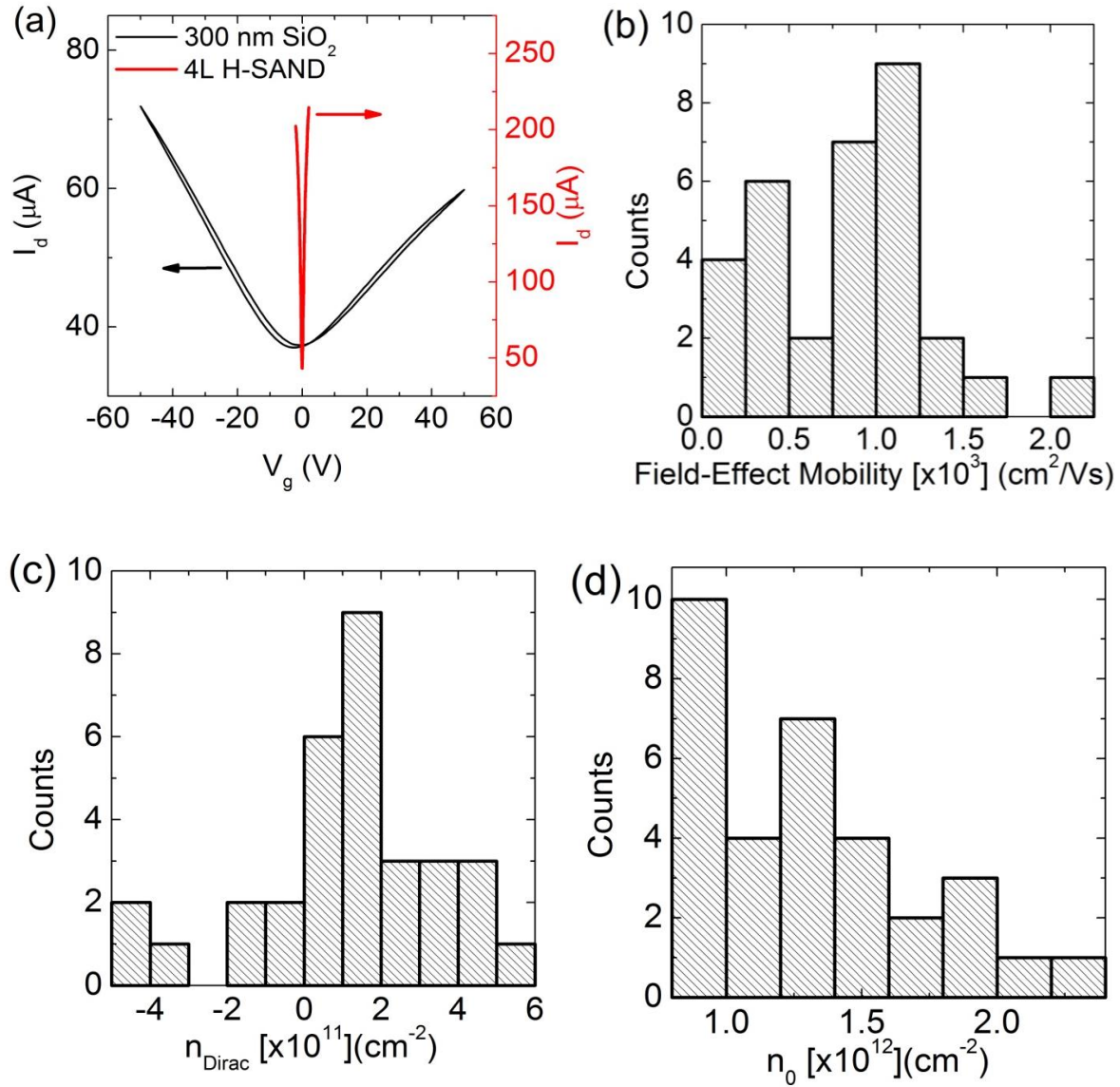

Figure S3. (a) A representative transfer characteristic of a control G-FET on a 300 nm $SiO_2$/Si substrate is compared with that of a G-FET on 4-layer Hf-SAND (from Fig. 2a, main text). Histograms of (b) field-effect mobility, (c) doping level ($n_{Dirac}$), and (d) residual carrier concentration at the Dirac point ($n_0$) extracted from fitting data from 32 control devices.

**Section S5: Controlled Doping of G-FETs by a SAM on Hf-SAND**

The smooth metal oxide capping layer of SAND allows functionalization with various self-assembled monolayers (SAMs) to tune electronic properties of semiconductors.[5] In particular, SAMs with a phosphonic acid linkage (similar to the linkage in the SAND PAE molecule) have been successfully demonstrated on Zr-SAND. Here, we choose an alkane-chain based hydrophobic octadecylphosphonic acid (ODPA) (see Fig. S4 for the chemical structure) as a prototype SAM to illustrate controlled doping of graphene on wafer-scale Hf-SAND. For this demonstration, a 3-layer SAND structure was fabricated on a 3-inch Si wafer by using ~2.5 nm thick $HfO_x$ interlayers compared to ~1 nm for the devices discussed in the main text.

Bottom-contact G-FETs were fabricated on an ODPA treated Hf-SAND wafer. The hydrophobic nature of the ODPA SAM precluded spin-coating of photoresist for photolithography. Therefore, the following processing was used to fabricate G-FETs. First, source-drain electrodes (2 nm Ti/ 70 nm Au) were patterned on the Hf-SAND wafer by photolithography and metal evaporation. The photoresist residues were cleaned from the Hf-SAND surface using brief RIE in $O_2$ plasma (power = 50 mW, $O_2$ flow-rate = 20 sccm, pressure = 100 mTorr, time = 10 sec). A quick test of contact angle of water on ODPA-treated Hf-SAND wafer revealed that this dry-etching step is critical to achieve successful self-assembly of ODPA on the post-lithography Hf-SAND surface. Then, the cleaned Hf-SAND wafers were kept in ODPA solution (2 mM solution in ethanol) for 12 hours in dark. Excess ODPA was removed from the surface *via* bath-sonication for 2 min. Then, CVD graphene was transferred from a Cu foil onto ODPA/Hf-SAND wafer following a procedure identical to the one used for Hf-SAND wafers. Note that the phosphonic linkage SAM is not expected to assemble on the Au contacts. Graphene channels were etched by photolithography and RIE. Although the SAM gets etched by RIE in the etched graphene regions of the wafer, the SAM under the graphene is protected by the photoresist mask. Finally, devices were thoroughly cleaned in photoresist stripper Remover PG. To characterize the leakage and capacitance of the ODPA/Hf-SAND structure, MIS capacitors were fabricated on the wafer after the deposition of ODPA and before the transfer of graphene. The overall increased thickness of the Hf-SAND results in a slightly lower leakage current density and almost 2x decreased capacitance compared to 4L Hf-SAND (Fig. S5).

Fig. S6a shows representative transfer characteristics of a G-FET on the ODPA/Hf-SAND wafer. Figs. S6b-d show histogram plots of device metrics extracted by fitting the data from a total of 111 measured devices with equation 1 in the main text. The distribution width of the device parameters of ODPA/Hf-SAND is comparable to that of a Hf-SAND wafer (Fig. 3, main text), suggesting that self-assembly of ODPA retains wafer-scale uniformity of the dielectric. The devices show substantial n-doping in graphene without appreciable compromise in mobility (Fig. S6c). Such n-doping by ODPA has also been previously reported in graphene devices on SAMs on oxidized aluminum dielectrics.[6]

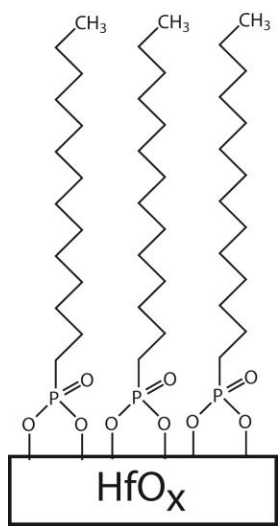

Figure S4. Chemical structure of an ODPA SAM on the capping layer (HfO$_x$) of Hf-SAND.

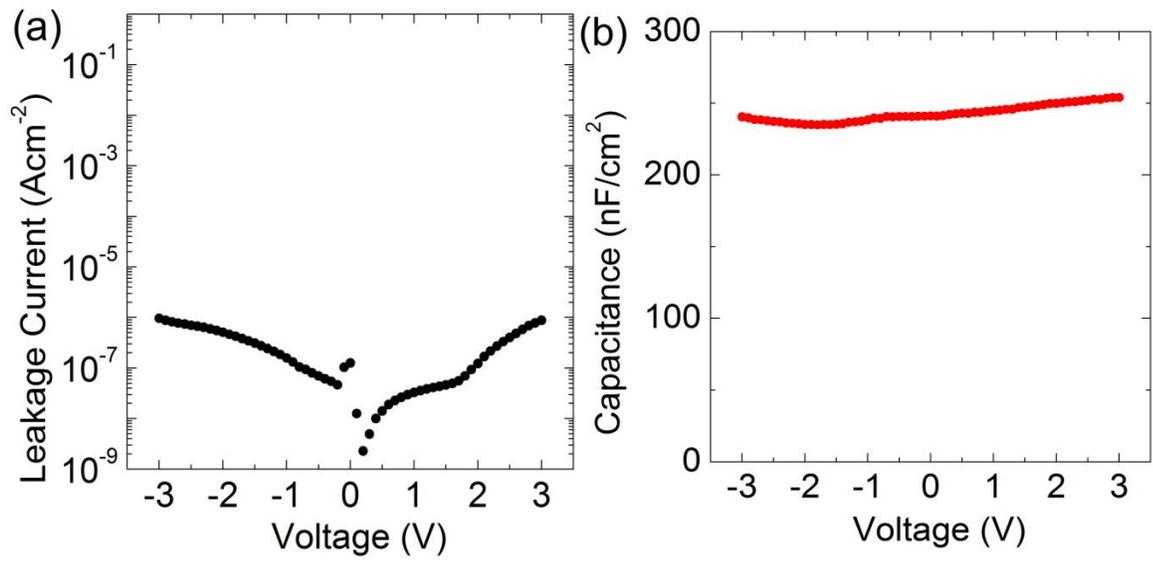

Figure S5. (a) Representative leakage current-voltage characteristics of a typical MIS capacitor on an ODPA/Hf-SAND wafer. (b) Capacitance-voltage characteristics of the same MIS capacitor taken at 10 kHz.

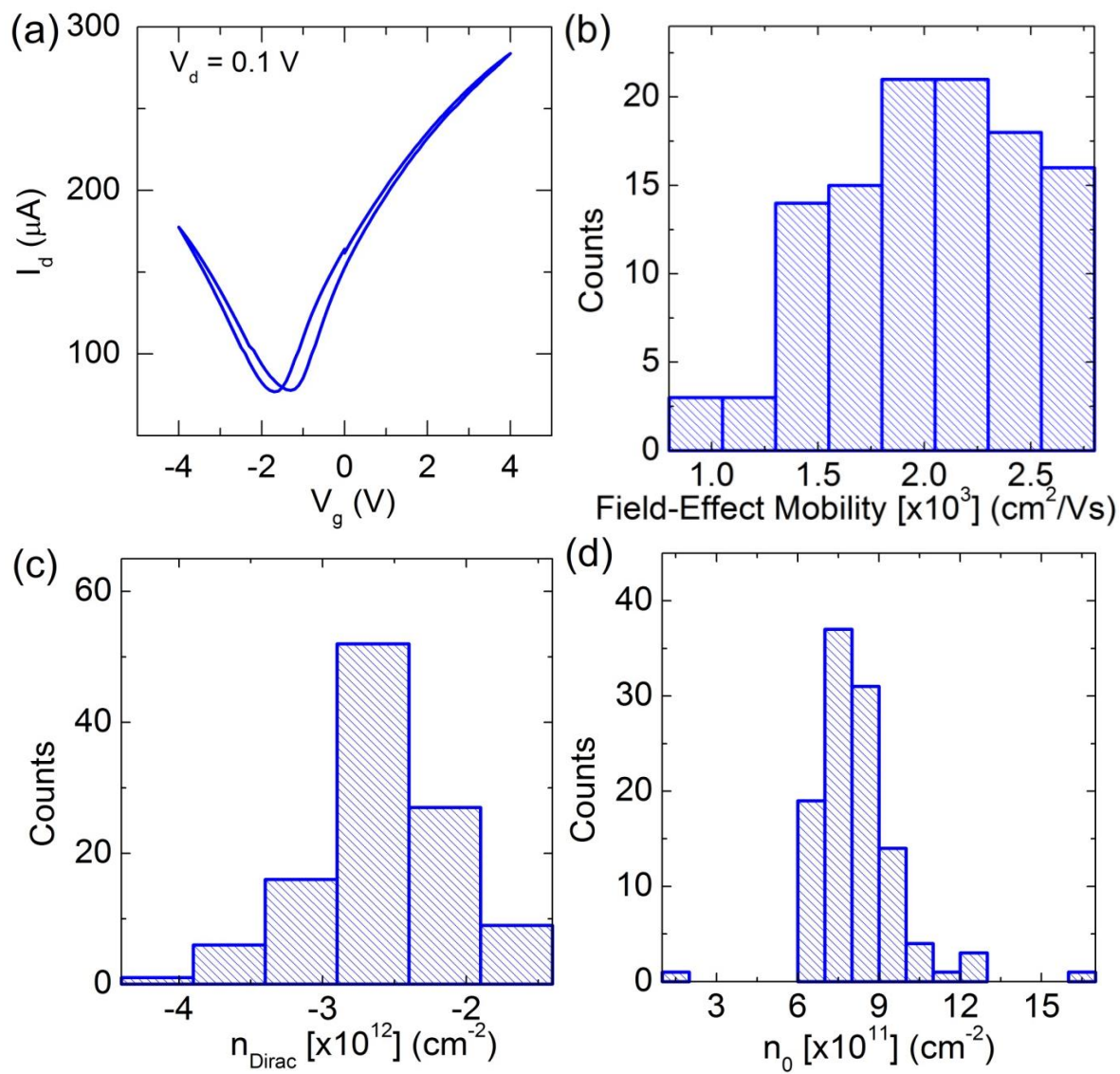

Figure S6. (a) Representative transfer characteristics of a G-FET ($L$ = 50 μm, $W$ = 150 μm) fabricated on ODPA-coated 3-layer Hf-SAND on a 3-inch wafer. Histograms of (b) hole mobilities, (c) $n_{Dirac}$, and (d) $n_0$ extracted from measurements of 111 devices.